# Lesion-Specific Prediction with Discriminator-Based Supervised Guided Attention Module Enabled GANs in Multiple Sclerosis

Jueqi Wang, Derek Berger, Erin Mazerolle, Jean-Alexis Delamer, Jacob Levman

Department of Computer Science, St Francis Xavier University, Canada
{x2019cwn, dberger, emazerol, jdelamer, jlevman}@stfx.ca

**Abstract.** Multiple Sclerosis (MS) is a chronic neurological condition characterized by the development of lesions in the white matter of the brain. $T_2$-fluid-attenuated inversion recovery (FLAIR) brain magnetic resonance imaging (MRI) provides superior visualization and characterization of MS lesions, relative to other MRI modalities. Follow-up brain FLAIR MRI in MS provides helpful information for clinicians towards monitoring disease progression. In this study, we propose a novel modification to generative adversarial networks (GANs) to predict future lesion-specific FLAIR MRI for MS at fixed time intervals. We use supervised guided attention and dilated convolutions in the discriminator, which supports making an informed prediction of whether the generated images are real or not based on attention to the lesion area, which in turn has potential to help improve the generator to predict the lesion area of future examinations more accurately. We compared our method to several baselines and one state-of-art CF-SAGAN model [1]. In conclusion, our results indicate that the proposed method achieves higher accuracy and reduces the standard deviation of the prediction errors in the lesion area compared with other models with similar overall performance.

**Keywords:** Image Synthesis, Longitudinal Prediction, Generative Adversarial Networks, Guided Attention, Multiple Sclerosis.

## 1 Introduction

Multiple sclerosis (MS) has been a leading cause of long-term neurological disability in young adults [2, 3]. MS is characterized by the development of lesions in the white matter (WM) of the brain. The progress of new lesions over time is an important criterion for MS diagnosis and treatment monitoring. Longitudinal magnetic resonance imaging (MRI) is an essential tool to identify the development of MS lesions over time [4]. Among different MRI sequences, $T_2$-fluid-attenuated inversion recovery (FLAIR) MRI provides improved visualization of periventricular MS lesions [5]. Thus, accurate prediction of follow-up FLAIR MRI scans could provide beneficial information for clinicians to monitor progression and designate treatment plans for MS patients. In this study, we propose a modification to generative adversarial net-works (GANs) to predict future lesion-specific FLAIR MRI for MS at fixed time intervals.

GANs have shown great potential for medical image synthesis [1, 3, 6-10]. Conventional GANs learn a global space mapping which focuses on whole brain changes and without specific focus on the comparatively small changes around regions-of-interest (ROIs), however, small alterations around lesion ROIs are more important clinically. To address this issue, medical image synthesis approaches that focus on ROIs specifically, have been developing recently [1, 3, 8-10].

Chen et al. [8] proposed a GAN based on a double input-output stream generator with a shared middle block which translates a whole image and a target area simultaneously, and uses two discriminators to perform global and ROI-specific mappings to enhance the quality of the predicted whole images and ROIs. They used unpaired data to do multi-modality translation between Computed Tomography, T1w and T2w MRI for organs (i.e. liver). However, this two-stream global and local mapping method may not be suitable for predicting future MS imaging. The location of MS lesions is inconsistent, and ROI sizes are particularly small compared to organs. Wei et al. [1] proposed a conditional flexible self-attention GAN (CF-SAGAN) for predicting positron emission tomography from MRI for MS using paired data. Their previous study [3] proposed a sketcher-refiner GAN architecture. The input images are required to go through two GANs successively. The first GAN generates preliminary anatomy and physiology information with a global loss function. The second GAN uses a weighted L1 loss to let the model pay more attention to lesion areas.

Despite the outstanding performance of existing techniques, these ROI-specific medical image synthesis methods focus on developing novel generator architectures [8, 9] and ROI-specific loss functions [1, 3, 10]. The attention of discriminators in GANs for ROI-specific medical image synthesis usually focuses on a global area and hasn't been shifted towards ROIs. In this paper, we propose to use 3D discriminator-based supervised pixel-level guided attention GAN (DGAGAN) to force the generator to generate images with high ROI accuracy, designed for predicting future MS MRIs, where the attention of the discriminator $D$ has been guided to focus on the ROI by incorporating pixel-level supervision in the discriminator $D$. Region-level supervised guided attention has been studied for medical image classification to incorporate additional pixel-level supervision to enhance the interpretation and accuracy of neural networks [11, 12, 22]. Zhong et al. [22] proposed to use a dilated and soft attention-guided convolutional neural network for breast cancer histology image classification. Unsupervised attention-augmented discriminators have been proposed to generate spatial attention maps from the discriminator to enhance the generator's capacity [13].

In this study, we use guided attention inference with supervision [14] in the discriminator to implement pixel-level supervised guided attention. Using guided attention in the discriminator supports the discriminator to locate and focus on the lesion area towards determining whether the presented image is real or not. Lesions can potentially change their appearance (i.e. grow spatially, go into remission, develop atrophy, etc.) anywhere across WM in MS patients' brains, and so, may benefit from guided attention to help automatically focus the learning machine towards prioritizing prediction accuracy in developing lesions and their immediate surrounding tissue.

This approach is unlike other unsupervised attention guided GAN methods [13, 15], which use the attention map provided by the discriminator to shift the generator's attention. Ground truth ROIs are provided during training in our study. The proposed DGAGAN uses a similar generator architecture as CF-SAGAN [1], which applies a weighted L1 loss to shift the generator's attention to MS lesions.

## 2 Methodology

### 2.1 Overview

We develop a deep learning model to predict follow-up $t_1$ timepoint MS longitudinal brain FLAIR images using previous $t_0$ timepoint MRI modalities at fixed time intervals into the future. GANs have become a popular solution to this medical image prediction application [6, 7]. There are usually two models in a typical GAN [16]: a generator $G$ and a discriminator $D$. The generator $G$ is designed to generate fake images $G(x)$ which could confuse the discriminator $D$ as much as possible. While the aim of the discriminator $D$, is to discern between real and fake images.

For discriminator enhanced ROI specific learning, the discriminator $D$ needs to have the capacity to localize the lesion location and to force its decisions to rely upon this lesion location. Most areas of the brain undergo subtle aging related changes, while the changes in appearance of MS lesions can be comparatively large and potentially of major clinical interest. Lesions in MS vary significantly in size, shape and image intensity [2]. The location of MS lesions is also inconsistent, and can present across white matter regions of the brain. The follow-up imaging lesion ROI could be different from the input time point (e.g. lesions might go into remission or become more prominent, new lesions may grow in the original ROI, and lesions may undergo atrophy). Due to such characteristics, in order to support the discriminator to locate the lesion and focus on the progressed lesion area during the prediction task, to determine if the given sample is real or fake, we apply a supervised guided attention [14] discriminator to guide the focus of the discriminator on the ROI from $t_1$ timepoint. We use a similar $G$ architecture as [1]. More details are provided in the specifications of the compared methods below.

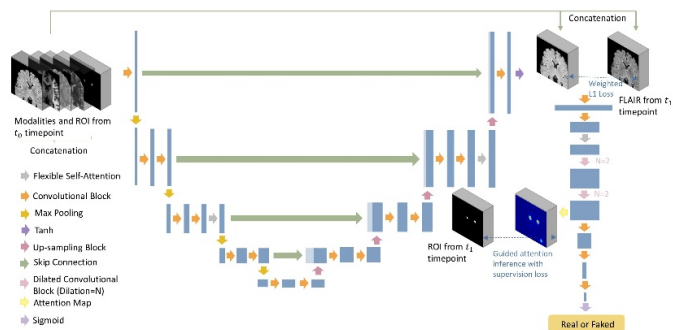

**Fig. 1.** Architecture for proposed discriminator-based guided attention GAN (DGAGAN).

## 2.2 Guided Attention Inference with Supervision

In order to incorporate pixel-level supervision in the discriminator $D$, we apply guided attention inference with supervision [14] in the discriminator $D$ to force the discriminator $D$ to make a decision as to whether this image is real or not based on the lesion areas from the $t_1$ timepoint.

Guided attention inference with supervision is capable of training the attention map from Grad-CAM [17]. Grad-CAM is a popular explanation method, which is capable of conveying the location of the spatial region upon which the given model bases its decisions / predictions. For a given image $I$, $f_{l,k}$ is the activation of unit $k$ in the $l$-th layer from the discriminator $D$, and $GAP(\cdot)$ is the global average pooling operation.

$$w_{l,k} = GAP\left(\frac{\partial s}{\partial f_{l,k}}\right) \tag{1}$$

$$A = ReLU\bigl(conv(f_l, w)\bigr) \tag{2}$$

where $l$ is the representation from the target dilated convolution layer (in Fig. 1), whose features contain both detailed spatial information and high-level semantics.

Then we use trilinear interpolation to upsample the attention map $A$ into the same dimensional space as the MS lesion ROI label $R_{les}$ to support the calculation of supervision loss $L_e$:

$$L_e = (Upsample(A) - R_{les})^2 \tag{3}$$

## 2.3 3D Discriminator-based Guided Attention GAN (DGAGAN)

In the original guided attention inference network [14], layer $l$ is the representation from the last convolutional layer, a reduced size feature map. However, medical images are generally high resolution and MS lesions are extremely small compared to the whole brain. This small feature map (from the last convolutional layer) typically contains degraded knowledge representation of the potential spatial information available, and thus may produce particularly poor knowledge representations of small lesions. Alternatively, layers before the layer $l$ also need to have enough parameters to accurately locate MS lesions, which is a difficult task, because of variable shape, size, position and contrast.

The task for layer $l$ is similar to a segmentation task. The output of layer $l$ is required to show roughly which pixels are lesion areas. Most of the discriminators in GANs use downsampling methods (e.g., max-pooling and striding) to enlarge the receptive field to capture high-level semantics. This has the effect of reducing the spatial resolution of the internal learned representation of the images [18]. In order to enlarge the receptive field without reducing the spatial resolution of feature maps, some semantic segmentation studies [18, 19] use dilated convolutions. In our study, we propose to use dilated convolutions with a guided attention inference network in the discriminator to support the discriminator to predict whether the image is real or not, while being informed by the lesion ROI. As shown in Fig. 1, following the exam-ple in [1], we use the self-attention layer [20] before the layer $l$ for its capacity to capture long-range dependencies across different lesion areas in WM.

Following the Pix2pix method [21], the discriminator $D$ is also conditioned on the input images $x$. The cross-entropy loss for the DGAGAN can be expressed as:

$$\mathcal{L}_{GAIGAN}(G, D) = \mathbb{E}_{x,y}[\log D(x, y)] + \mathbb{E}_x\left[\log\left(1 - D(x, G(x))\right)\right] \quad (4)$$

where $\mathbb{E}$ is the maximum likelihood estimation, $x, y$ denotes the input images and $x$'s corresponding follow-up target image, separately. The final objective for DGAGAN is:

$$\arg\min_G \max_D \mathcal{L}_{GAIGAN}(G, D) + \lambda_{l1} L_{l1}(G) + \omega L_e \quad (5)$$

where $\lambda_{l1}$ is a non-negative trade-off parameter, which is used to balance the adversarial and the normalized weighted L1 loss. The normalized weighted L1 loss $L_{l1}(G)$ is expressed in section 3.3. $\omega$ is the weighting parameter for guided attention with supervision loss.

## 3   Materials

### 3.1   Dataset

To validate our method, we used the ISBI2015 longitudinal MS dataset [26]. Five participants, for which the dataset provides corresponding lesion ROIs are used in this study. Four participants had scans at four time points, and one participant had scans at five time points. Each scan was acquired on the same MRI scanner. The first time-point MPRAGE was rigidly registered into 1 mm isotropic MNI template space and used as a baseline for the remaining images from the input and follow-up time-points. T1-w MPRAGE, T2-w, PD-w, and FLAIR modalities are provided for each time point. Consecutive time-points of all participants are approximately one year apart. Our models predict follow-up images at one year in advance, as such 16 samples are available in this dataset.

### 3.2   Evaluation metrics

There are two kinds of image quality that should be considered for image synthesis performance: intensity and perceptual distance between the ground-truth image and its predicted image. In order to take both qualities into account, two popular metrics are used in this study: peak signal-to-noise ratio (PSNR), and structural similarity index (SSIM) [23]. SSIM and PSNR are provided for the whole image, while only PSNR is provided on the target $t_1$ timepoint ROI.

### 3.3 Compared methods

**3D U-Net.** A 6-level 3D U-Net is utilized as a baseline model and the generator for cGAN. The associated ROI is concatenated with all modalities from the $t_0$ timepoint along the channel dimension into the 3D U-Net to provide the model with lesion information at the early timepoint. L1 distance is used as a loss function.

**3D Conditional GAN (3D cGAN).** This method is an extended version of the Pix2pix [21] method into 3D. The input images and the architecture of the generator $G$ is the same as the 3D U-Net. All the input images of the generator are concatenated with real images $y$ or fake images $G(x)$ and used as input in the discriminator $D$.

**CF-SAGAN** [1]**.** The architecture of CF-SAGAN[1] is the same as the proposed DGAGAN as shown in Fig. 1. The generator of CF-SAGAN uses two flexible self-attention layers, which the 3D U-Net and the 3D cGAN do not use. The input of the generator $G$ and discriminator $D$ in CF-SAGAN is the same as the 3D cGAN. The similar normalized weighted L1 loss[2] is used as in [1]:

$$L_{l1}(G) = \frac{1}{2m}\sum_{i=1}^{m} \omega_i \left|y^i - G(x)^i\right|, \omega_i = \begin{cases} 1 - \frac{m_{Les}}{m}, i \in R_{Les} \\ 1 - \frac{m_{WM}}{m}, i \in R_{WM} \\ 1 - \frac{m_{other}}{m}, i \in R_{other} \end{cases} \quad (6)$$

Where $R_{Les}$ is the ROI from the $t_1$ time point, $R_{WM}$ is the WM area outside the lesion area, and $R_{other}$ is the remaining area beyond these two ROIs, with the number of voxels in each region $m_{Les}, m_{WM}, m_{other}$.

### 3.4 Implementation Details

An image size of (150, 190, 150) was cropped out to reduce the background region and save computational resources. Each modality volume was linearly scaled to [-1, 1] from the original intensity for normalization. To fit the 3D image into the generator and make the whole model fit into GPU memory, we split them into eight overlapping patches of size (128, 128, 128). The overlapped regions are averaged to aggregate those patches. Because of our small dataset, data augmentation is necessary for the invariance and robustness properties of model performance [24]. A data augmentation of rotation with a random angle $[-18°, 18°]$, a random spatial scaling factor $[0.85, 1.15]$, and a random flip in three axes was employed for all the methods during training. Batch size of 1 was used for all the methods. 5-fold cross validation was applied at the participant level. To balance the generator and discriminator in GANs, we use label smoothing to improve the stability of training the GANs. The WM masks used in CF-SAGAN and our proposed DGAGAN are generated from [25].

---

[1] A sketcher-refiner architecture has been used in the original CF-SAGAN paper, while we did not implement it because our task involves a previous round of FLAIR imaging, resulting in very similar global image properties in the input and target.

[2] In [1], their normalization weighted L1 loss is averaged across patients to alleviate the influence of MS lesion size variability. However, their normalization weighted L1 loss didn't work on our problem. Computing the weights on each patch produces superior performance, potentially because of our small sample size.

Network parameters were optimized through Adam with momentum parameters $\beta_1 = 0.5$ and $\beta_2 = 0.999$. Weight decay $\lambda = 7 \times 10^{-8}$ and $\lambda = 1 \times 10^{-5}$ is used for the generator $G$ and the discriminator $D$, respectively. For all models, we conduct 300 epochs of training. The learning rate is fixed as $2 \times 10^{-4}$ and $1 \times 10^{-5}$ for the generator and discriminator, respectively. For a fair comparison, the same hyper-parameters were employed across all the architectures. The number of parameters of the proposed discriminator is fewer than other compared methods, while the number of parameters for the generator and the U-Net are similar. Detailed information for the number of parameters can be found in Table 1 in the supplementary materials.

$\lambda_{l1}$ is set to 300 for 3D cGAN, while $\lambda_{l1}$ is set to 1800 for CF-SAGAN and our proposed DGAGAN because of the usage of normalized weighted L1 loss. $\omega$ is set to 0 during the first 100 epochs, and linearly increases from 0 to 10 in the following 100 epochs, and then stays at 10 for the next 100 epochs.

## 4 Results and Discussion

Table 1 shows the quantitative results obtained by different methods regarding mean PSNR, SSIM and their standard deviation for the whole image, as well as PSNR for the MS lesion region / ROI from $t_1$. The aggregated 3D volume is used to compute the metrics in order to represent the performance on the whole scan. Each volume was linearly scaled to $[0, 1]$ to compute the metrics to guarantee fair comparison. Firstly, we can observe that the 3D UNet with a fixed L1 loss function achieves better results than the 3D cGAN on SSIM of the whole image and PSNR of the lesion ROI. This could potentially mean that the fixed overall L1 loss function is better than a simple discriminator balanced with L1 loss to capture small changes for future MRI image prediction, since the target $t_1$ FLAIR is very similar to the input $t_0$ FLAIR image, in 1 year apart. Additionally, a performance improvement with less standard deviation has been demonstrated in the ROI when we use our proposed DGAGAN method compared to other methods. DGAGAN also achieved similar performance on the whole image metric compared with alternative methods, which may indicate that the proposed DGAGAN method might be a good way to apply GAN methods in longitudinal medical image prediction. In this situation, some subtle aging occurs in a large area of the brain, while relatively large changes happen around some small lesion areas. Although we use a small dataset with only 5 participants and 16 samples, thanks to data augmentation, K-Fold cross validation, and splitting our volumes into overlapping patches, we still achieve promising results for predicting future MS images.

**Table 1.** Quantitative Evaluation Results of Different Methods (mean ± standard deviation), obtained by evaluated methods on the validation folds.

| Methods | Whole image | | MS lesion region |
|---|---|---|---|
| | PSNR ↑ | SSIM ↑ | PSNR ↑ |
| 3D UNet | 28.81±2.276 | **0.8849±0.026** | 19.58±3.398 |
| 3D cGAN | 29.33±2.148 | 0.8678±0.026 | 18.84±1.627 |
| CF-SAGAN | **29.79±1.766** | 0.8830±0.022 | 19.49±1.793 |
| DGAGAN | 29.74±1.608 | 0.8834±0.023 | **20.05±1.425** |

In Fig. 2, the enlarged area exhibits a hypointensity in the center of the lesion on the target image, which is best predicted by the proposed DGAGAN, which implies the technique is the superior method among those compared for modeling longitudinal lesion changes. The second row of the figure is the attention map from the discriminator of the GAN using the real image as input. It is noteworthy that the attention map in the discriminator of the proposed DGAGAN approach successfully locates the ROI.

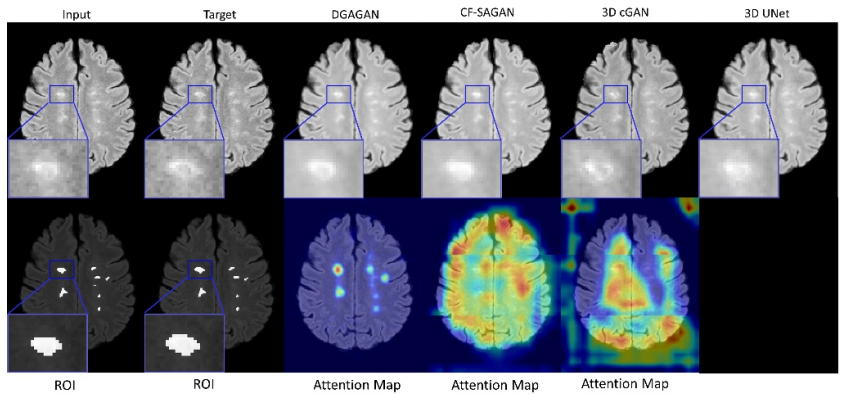

**Fig. 2.** Example predicted images in the validation fold are demonstrated in the first row. The second row of the figure shows the lesion areas from the corresponding timepoint and attention maps from the discriminator of the respective GANs.

In conclusion, we extended how to enhance the performance of predicting longitudinal MS lesion changes through GANs by exploring the architecture of the discriminator. We propose DGAGAN to use guided attention inference with supervision [14] and dilated convolutions in the discriminator to generate better ROIs, especially for prediction of future MS MRI examinations with a focus on lesions specifically.

Future work will incorporate the ROI based approaches with attention modules presented in this manuscript with methods for modeling time as a continuous spatially distributed variable, helping the techniques simultaneously focus on ROIs corresponding to lesions while modeling spatially variable developmental trajectories to assist in modeling lesion progression, atrophy and remission.

# References

1. Wei, W., Poirion, E., Bodini, B., Tonietto, M., Durrleman, S., Colliot, O., Stankoff, B., Ayache, N.: Predicting PET-derived myelin content from multisequence MRI for individual longitudinal analysis in multiple sclerosis. NeuroImage 223, 117308 (2020)
2. Zhang, H., Zhang, J., Li, C., Sweeney, E.M., Spincemaille, P., Nguyen, T.D., Gauthier, S.A., Wang, Y., Marcille, M.: ALL-Net: Anatomical information lesion-wise loss function integrated into neural network for multiple sclerosis lesion segmentation. NeuroImage: Clinical 32, 102854 (2021)
3. Wei, W., Poirion, E., Bodini, B., Durrleman, S., Ayache, N., Stankoff, B., Colliot, O.: Predicting PET-derived demyelination from multimodal MRI using sketcher-refiner adversarial training for multiple sclerosis. Medical Image Analysis 58, 101546 (2019)
4. McGinley, M.P., Goldschmidt, C.H., Rae-Grant, A.D.: Diagnosis and Treatment of Multiple Sclerosis: A Review. JAMA 325, 765-779 (2021)
5. Wei, W., Poirion, É., Bodini, B., Durrleman, S., Colliot, O., Stankoff, B., Ayache, N.: Fluid-attenuated inversion recovery MRI synthesis from multisequence MRI using three-dimensional fully convolutional networks for multiple sclerosis. Journal of Medical Imaging 6, 1 (2019)
6. Wegmayr, V., Hörold, M., Buhmann, J.M.: Generative Aging Of Brain MRI For Early Prediction Of MCI-AD Conversion. In: 2019 IEEE 16th International Symposium on Biomedical Imaging (ISBI 2019), pp. 1042-1046. (2019)
7. Xia, T., Chartsias, A., Wang, C., Tsaftaris, S.A.: Learning to synthesise the ageing brain without longitudinal data. Medical Image Analysis 73, 102169 (2021)
8. Chen, J., Wei, J., Li, R.: Targan: Target-aware generative adversarial networks for multi-modality medical image translation. pp. 24-33. Springer,  (2021)
9. Yu, B., Zhou, L., Wang, L., Shi, Y., Fripp, J., Bourgeat, P.: Sample-Adaptive GANs: Linking Global and Local Mappings for Cross-Modality MR Image Synthesis. IEEE Transactions on Medical Imaging 39, 2339-2350 (2020)
10. Finck, T., Li, H., Schlaeger, S., Grundl, L., Sollmann, N., Bender, B., Bürkle, E., Zimmer, C., Kirschke, J., Menze, B., Mühlau, M., Wiestler, B.: Uncertainty-Aware and Lesion-Specific Image Synthesis in Multiple Sclerosis Magnetic Resonance Imaging: A Multicentric Validation Study. Front Neurosci 16, 889808 (2022)
11. Son, J., Bae, W., Kim, S., Park, S.J., Jung, K.-H.: Classification of Findings with Localized Lesions in Fundoscopic Images Using a Regionally Guided CNN. In: Computational Pathology and Ophthalmic Medical Image Analysis, pp. 176-184. Springer International Publishing,  (2018)
12. Yang, H., Kim, J.-Y., Kim, H., Adhikari, S.P.: Guided soft attention network for classification of breast cancer histopathology images. IEEE transactions on medical imaging 39, 1306-1315 (2019)
13. Lin, Y., Wang, Y., Li, Y., Gao, Y., Wang, Z., Khan, L.: Attention-Based Spatial Guidance for Image-to-Image Translation. In: 2021 IEEE Winter Conference on Applications of Computer Vision (WACV), pp. 816-825.  (2021)
14. Li, K., Wu, Z., Peng, K.C., Ernst, J., Fu, Y.: Guided Attention Inference Network. IEEE Transactions on Pattern Analysis and Machine Intelligence 42, 2996-3010 (2020)
15. Emami, H., Dong, M., Glide-Hurst, C.K.: Attention-Guided Generative Adversarial Network to Address Atypical Anatomy in Synthetic CT Generation. In: 2020 IEEE 21st

International Conference on Information Reuse and Integration for Data Science (IRI), pp. 188-193. (2020)

16. Goodfellow, I., Pouget-Abadie, J., Mirza, M., Xu, B., Warde-Farley, D., Ozair, S., Courville, A., Bengio, Y.: Generative adversarial nets. Advances in neural information processing systems 27, (2014)
17. Selvaraju, R.R., Cogswell, M., Das, A., Vedantam, R., Parikh, D., Batra, D.: Grad-CAM: Visual Explanations from Deep Networks via Gradient-Based Localization. In: 2017 IEEE International Conference on Computer Vision (ICCV), pp. 618-626. (2017)
18. Li, W., Wang, G., Fidon, L., Ourselin, S., Cardoso, M.J., Vercauteren, T.: On the Compactness, Efficiency, and Representation of 3D Convolutional Networks: Brain Parcellation as a Pretext Task. In: Information Processing in Medical Imaging, pp. 348-360. Springer International Publishing, (2017)
19. Chen, L.-C., Papandreou, G., Kokkinos, I., Murphy, K., Yuille, A.L.: Deeplab: Semantic image segmentation with deep convolutional nets, atrous convolution, and fully connected crfs. IEEE transactions on pattern analysis and machine intelligence 40, 834-848 (2017)
20. Zhang, H., Goodfellow, I., Metaxas, D., Odena, A.: Self-Attention Generative Adversarial Networks. In: Kamalika, C., Ruslan, S. (eds.) Proceedings of the 36th International Conference on Machine Learning, vol. 97, pp. 7354--7363. PMLR, Proceedings of Machine Learning Research (2019)
21. Isola, P., Zhu, J.-Y., Zhou, T., Efros, A.A.: Image-to-image translation with conditional adversarial networks. In: Proceedings of the IEEE conference on computer vision and pattern recognition, pp. 1125-1134. (2017)
22. Zhong, Y., Piao, Y., Zhang, G.: Dilated and soft attention-guided convolutional neural network for breast cancer histology images classification. Microsc Res Tech 85, 1248-1257 (2022)
23. Zhou, W., Bovik, A.C., Sheikh, H.R., Simoncelli, E.P.: Image quality assessment: from error visibility to structural similarity. IEEE Transactions on Image Processing 13, 600-612 (2004)
24. Ronneberger, O., Fischer, P., Brox, T.: U-Net: Convolutional Networks for Biomedical Image Segmentation. In: Medical Image Computing and Computer-Assisted Intervention – MICCAI 2015, pp. 234-241. Springer International Publishing, (2015)
25. Cerri, S., Hoopes, A., Greve, D.N., Mühlau, M., Van Leemput, K.: A longitudinal method for simultaneous whole-brain and lesion segmentation in multiple sclerosis. Machine Learning in Clinical Neuroimaging and Radiogenomics in Neuro-oncology, pp. 119-128. Springer (2020)
26. Carass, A., Roy, S., Jog, A., Cuzzocreo, J.L., Magrath, E., Gherman, A., Button, J., Nguyen, J., Prados, F., Sudre, C.H., Jorge Cardoso, M., Cawley, N., Ciccarelli, O., WheelerKingshott, C.A.M., Ourselin, S., Catanese, L., Deshpande, H., Maurel, P., Commowick, O., Barillot, C., Tomas-Fernandez, X., Warfield, S.K., Vaidya, S., Chunduru, A., Muthuganapathy, R., Krishnamurthi, G., Jesson, A., Arbel, T., Maier, O., Handels, H., Iheme, L.O., Unay, D., Jain, S., Sima, D.M., Smeets, D., Ghafoorian, M., Platel, B., Birenbaum, A., Greenspan, H., Bazin, P.-L., Calabresi, P.A., Crainiceanu, C.M., Ellingsen, L.M., Reich, D.S., Prince, J.L., Pham, D.L.: Longitudinal multiple sclerosis lesion segmentation: Resource and challenge. Neuroimage. 148, 77–102 (2017).